\documentclass[a4paper, amsfonts, amssymb, amsmath, reprint, showkeys, nofootinbib, twoside,superscriptaddress, aps, prb,longbibliography]{revtex4-2}
\usepackage[english]{babel}
\usepackage[utf8]{inputenc}
\usepackage[colorinlistoftodos, color=green!40, prependcaption]{todonotes}
\usepackage{amsthm}
\usepackage{mathtools, bm, comment}
\usepackage{physics}
\usepackage{xcolor}
\usepackage{graphicx}
\usepackage[left=23mm,right=13mm,top=35mm,columnsep=15pt]{geometry} 
\usepackage{adjustbox}
\usepackage{placeins}
\usepackage[T1]{fontenc}
\usepackage{lipsum}
\usepackage{csquotes}
\usepackage[caption=false]{subfig}
\usepackage{subfig}
\usepackage{multirow}

\begin{document}
\title{Drivers of chemical diffusion of hydrogen in the thin transition metallic glass V$_{80}$Zr$_{20}$}

\author{Lennart Spode}
\affiliation{Division of Materials Physics, Department of Physics and Astronomy, Uppsala University, Box 516, SE-75121, Uppsala, Sweden}
\author{Ola Hartmann}
\affiliation{Division of Materials Physics, Department of Physics and Astronomy, Uppsala University, Box 516, SE-75121, Uppsala, Sweden}

\author{Gunnar K. P\'alsson}
    \email[Correspondence email address: ]{gunnar.palsson@physics.uu.se}
    \affiliation{Division of Materials Physics, Department of Physics and Astronomy, Uppsala University, Box 516, SE-75121, Uppsala, Sweden}
\date{\today} 

\begin{abstract}\noindent
 We demonstrate the feasibility of using optical transmission to determine concentration-dependent hydrogen diffusion coefficients and activation energies of thin metallic glass films over a wide range of temperatures and concentrations. The hydrogen concentration's temporal and spatial profiles are simultaneously extracted without requiring a metal-insulator transition or hydride formation. The concentration-dependent activation energy is extracted and is found to exhibit similar concentration dependence as hydrogen in the disordered bulk $\alpha$-phase of vanadium. The activation energy is in stark contrast, however, to the activation energy found in epitaxial vanadium films of similar thickness. The chemical diffusion increases with hydrogen concentration in the glass, whereas it decreases in the crystalline case owing to the differences in the thermodynamic factors. This technique provides detailed insight into the concentration dependence of hydrogen diffusion and constrains theories treating correlated motion of hydrogen and short-range hydrogen-hydrogen interaction. The method can be applied to other materials that show an optical response to hydrogen uptake.
\end{abstract}
 
\keywords{thin films, hydrogen, amorphous, metallic glass, hydrogen diffusion, VZr, optical transmission, hydrogen-hydrogen interaction, correlation factor}

\maketitle
\section{Introduction} \label{sec:introduction}
Hydrogen interactions with metals are intriguing from both a fundamental \cite{10.1023/a:1008748627295, 10.1016/j.jallcom.2023.168945, 10.1039/c1fd00105a} and an applied point of view such as for hydrogen storage and material durability~\cite{10.3389/fenrg.2016.00011}. Specifically, hydrogen diffusion in disordered metals exhibits experimental anomalies such as non-Arrhenian diffusion~\cite{Eliaz1999a,Eliaz1999b}. Kirchheim, Richards and others have developed theoretical models for describing the motion of hydrogen in metallic glasses \cite{10.1016/0001-6160(82)90002-5, 10.1103/physrevb.27.2059, 10.1088/0022-3719/20/10/013}, however, many aspects of the mechanisms that drive it are still unknown, like the influence of the coupling to a thermal bath~\cite{10.1103/physrevlett.60.2307}, the dynamics of self-trapping, the appropriateness of the description of hydrogen as a small polaron quasi particle~\cite{10.1103/physrevb.26.6455,10.1103/physrevlett.105.185901}, or what drives the non-Arrhenius-like temperature dependence of the diffusion coefficient $D$. Furthermore, the potential role of topological disorder, mobility correlation, short-range hydrogen-hydrogen interaction and its effects on blocking and site-availability, and the fractal character of the network of diffusive paths remain poorly understood. This is partly due to the difficulty of accurately modeling the influence of the amorphous host structure and partly due to a lack of experimental data required to test the different models. To validate candidate models, it is highly desirable to have consistent data over a wide range of both concentrations and temperatures for both the diffusion coefficient and the activation energy.

The diffusion of the absorbed hydrogen has generally been described with over-barrier hopping between local interstitial sites~\cite{Fukai2005H,10.1103/physrevlett.105.185901}. However, the motion of hydrogen in metals exhibits a plethora of other interesting phenomena, ranging from quantum tunneling and Kondo physics to critical slowing down \cite{Fukai2005H,10.1007/bfb0103398}. A strain field emerges during the dynamic self-trapping process, which together with the charge, can be viewed as a small polaron~\cite{10.1016/0001-6160(82)90002-5,10.1088/0022-3719/20/10/013}, a quasi-particle with peculiar properties. Ab-initio molecular dynamics calculations have even shown that hydrogen can decouple from its strain field, bringing about a new range of diffusive behavior \cite{10.1103/physrevlett.105.185901}.

Thin films offer an interesting playground to investigate metallic glasses due to the high achievable quenching rates during their growth, making it possible to grow films with a high degree of uniformity while avoiding inclusions, voids and other defects \cite{10.1016/j.nocx.2021.100061}. They allow well-controlled composition, surface termination \cite{Wang2008, 10.1016/j.jallcom.2009.12.178}, and material interfaces \cite{10.1016/j.ijhydene.2014.08.035}. Furthermore, metallic glasses can be realized with compositions not accessible with bulk methods \cite{Kaplan2022}, offering access to unexplored territory. These meta-stable phases can occur in regions forbidden in the corresponding crystalline phase diagram~\cite{Kaplan2022}. However, measuring the diffusion coefficient in thin films is currently not possible using standard techniques such as the Gorsky effect, nuclear magnetic resonance, or quasi-elastic neutron scattering due to the low sample volume.

In 2012, a method was demonstrated that allows to determine hydrogen diffusion in crystalline vanadium films optically, by relying on the linear dependence between optical transmission and hydrogen concentration~\cite{Palsson2012}. 
A great advantage is that both the temporal and spatial profiles are obtained simultaneously, thereby providing more accurate diffusion coefficients. This enabled a number of studies using the technique to look at diffusion in artificially made structures such as superlattices~\cite{10.1103/physrevb.95.064310, Palsson2012,Huang2016, Bylin2024}. This method furthermore, circumvents the need for using indicator layers~\cite{PhysRevB.66.020101}, which provide indirect measurements of hydrogen concentration. Another advantage of the technique is that the concentration dependence over a wide range can be determined from a single measurement, making it unnecessary to use many different samples with different hydrogen concentrations. \\

The band structure of a metallic glass is not a well-defined concept, and inter-band transitions that drive optical absorption are thus present in a wider energy range~\cite{Bylin2024}. Therefore, it needs to be shown whether the method can be applied to disordered materials. In the following, we demonstrate the feasibility of using this contactless, non-destructive, and indicator-free method to determine diffusion coefficients as a function of concentration to study the diffusion of hydrogen in metallic glasses. 

\section{Methods} \label{sec:methods}
\subsection{Sample design and growth}
\begin{figure}[h!]
    \centering
    \includegraphics[width = \linewidth]{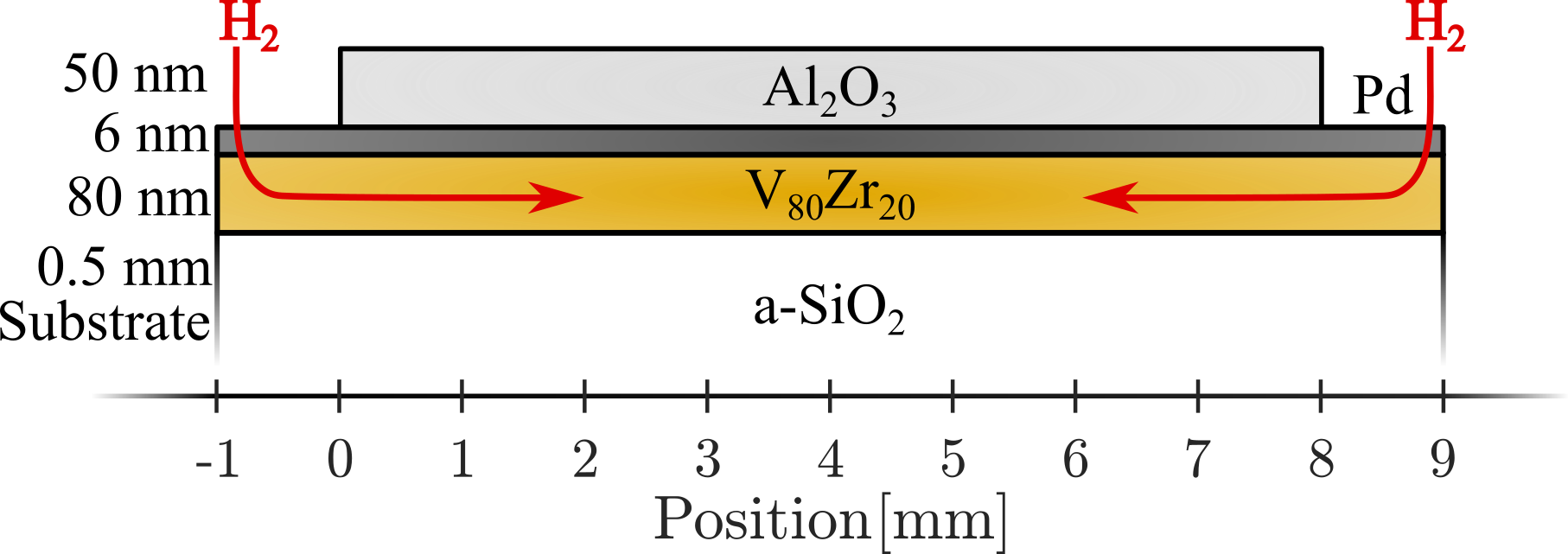}
    \caption{The film consists of $80\,$nm of V$_{0.8}$Zr$_{0.2}$, grown under UHV conditions onto $10\,$mm$\times 10\,$mm$\times 0.5\,$mm a-SiO$_2$ glass using magnetron sputtering. The sample is covered by $6\,$nm of palladium and $50\,$nm of Al$_2$O$_3$, enabling hydrogen uptake only through two opposite stripes of exposed palladium.}
    \label{fig:sampledesign}
\end{figure}

The samples were produced using DC and AC magnetron sputtering. The design is shown in Fig. \ref{fig:sampledesign}. Each sample consisted of $80\,$nm of V$_{0.8}$Zr$_{0.2}$, grown onto a $10\,$mm$\times 10\,$mm$\times 0.5\,$mm double-side polished a-SiO$_2$ substrate. The film was covered with $6\,$nm of palladium and the center $8\,$mm$\times 10\,$mm were subsequently covered using a mask to allow a $50\,$nm thick Al$_2$O$_3$ layer to be grown. Vanadium, zirconium and Al$_2$O$_3$ were grown in a chamber having base pressure of $1.5\times 10^{-10}\,$mbar and argon pressure of $3.33\times10^{-3}\,$mbar, whereas the palladium layer was grown in an adjacent chamber with a base pressure of $5\times 10^{-8}\,$mbar and argon pressure of $6.66\times10^{-3}\,$mbar. The power-dependent growth rate was calibrated for each target to determine the required sputtering times and power ratios for the combinatorial sputtering process. The sample thicknesses were determined using x-ray reflectometry and extracted by fitting a slab model to the reflectivity data using Gen-X \cite{Björk2007, Björk2011}. The substrates were baked for $50$ minutes at $350\,^\circ$C, to reduce possible deposits of water and organic contaminants. During growth, the temperature was decreased to $27\,^\circ$C to ensure maximally amorphous sample growth \cite{10.1016/0022-5088(84)90091-2, 10.1107/s1600576724006368}. Sample composition and thickness for the V-Zr system have been verified previously using X-ray reflectometry, X-ray diffraction and Rutherford backscattering \cite{Bylin2024, Kaplan2022}, verifying the growth procedure to gain high-quality, X-ray amorphous samples. The palladium layer protects the underlying V$_{0.8}$Zr$_{0.2}$ from oxidation and catalyzes the hydrogen dissociation without taking up hydrogen in the pressure and temperature range applied \cite{10.1007/3-540-08883-0}. Al$_2$O$_3$ is inert towards reactions with hydrogen \cite{Wang2008} and, therefore, acts as a barrier, allowing hydrogen uptake into the sample only through the free surface of the catalytic palladium layer. The sample design facilitates diffusion of hydrogen only towards the center of the sample, as shown in Fig. \ref{fig:sampledesign}, which is important for determining the diffusion coefficients. A total of 6 identical samples were grown in two separate growth sessions. Each sample was subjected to a hydrogen atmosphere and temperature corresponding to a nominal maximum hydrogen concentration of $0.7\,$H/M. This was to ensure that the hydrogen diffusion was measured corresponding to a well-defined as-grown configuration of the metal atoms, as the influence of hydrogen cycling is not yet known in V$_{0.8}$Zr$_{0.2}$ metallic glass thin films. This choice is not necessary to successfully use the method.
\subsection{Measurement technique}
The optical technique used to determine the hydrogen diffusion in crystalline samples is explained in detail in \cite{Palsson2012} and \cite{Huang2016}. Briefly, the samples are placed in a vacuum chamber with two windows: one above and one below the sample. The sample needs to be thin enough to allow for appreciable light transmission but thick enough to show a measurable signal when absorbing hydrogen. A monochromatic, incoherent light-emitting diode is used to minimize interference from the front and back sides of the windows. The sample is exposed to a temperature through external heating of the chamber and a hydrogen atmosphere of pressure $p$. The atomic and electronic structures of the sample change when hydrogen is absorbed \cite{Bylin2024}, which can cause a change in optical transmission. The changes in transmission are registered through a lens-camera system the focal point of which lies in the film plane. In the present experiments, a red diode with wavelength $\lambda = 625\,$nm was used. Images were recorded every $30\,$s to $60\,$s. The hydrogen evolution was tracked over a distance of $2.5\,$mm to $3\,$mm from one edge of the palladium/Al$_2$O$_3$ interface. The applied hydrogen pressure at each temperature is shown in Tab. \ref{tab:hpressure}. These pressures were selected to give approximately the same final concentration as determined from \cite{Bylin2024,Bylin2022}. The base pressure of the vacuum chamber was at $1\times10^{-8}\,$mbar, with the residual gas dominated by hydrogen. The hydrogen was supplied by a gas bottle of 6N purity that was subsequently passed through a Nupure purifier, bringing the hydrogen purity to the ppb level. \\

\begin{table}[h!]
    \centering
    \begin{tabular}{c|c}
        Temperature [K] & Loading pressure [mbar] \\
    \hline
        $323$ &  $2.47$ \\ 
        $348$ &  $7.31$\\
        $373$ & $20.6$ \\ 
        $398$ & $58.7$ \\ 
        $423$ & $169$ \\ 
        $448$ &  $488$   
    \end{tabular}
    \caption{Applied hydrogen pressure at each temperature.}
    \label{tab:hpressure}
\end{table}

Spatial and temporal transmission profiles are extracted from the images. The transmission profiles were converted into hydrogen concentration profiles $c(\textbf{x},t)$ using the calibration curve in [Ref.~\cite{Bylin2024}].
\subsection{Theoretical background}
\label{sec:theory}

The hydrogen diffusion inside the samples can be seen as a one-dimensional process due to the symmetry of the sample. If the concentration profile is known, the diffusion coefficient can be extracted using Fick's second law of diffusion~\cite{Fick1855, Matano1933}:

\begin{equation}
    \frac{\partial c(x,t)}{\partial t} =  \frac{\partial}{\partial x} \left( D(c,T) \cdot \frac{\partial c(x,t)}{\partial x}\right),
    \label{eq:diff} 
\end{equation}
where $D(c,T)$ is the concentration-dependent (and temperature-dependent) diffusion coefficient. The diffusion coefficient can be extracted using the Stenlund method~\cite{Stenlund2011}, which yields:
    
\begin{equation}
    D(c,T)=\frac{-\int_x^{\infty}\frac{\partial c\left(x_1, t\right)}{\partial t} \cdot d x_1}{\frac{\partial c(x, t)}{\partial x}} 
    \label{eq:stenlund}
\end{equation}
    
It is assumed that the following boundary condition is true: 
\begin{equation}   
\lim _{x \rightarrow \infty} \frac{\partial c(x, t)}{\partial x} = 0.
\label{eq:stenlundcondition}
\end{equation}
 The derivatives $\frac{\partial c(x, t)}{\partial x}$ and $\frac{\partial c(x, t)}{\partial t}$ as well as the integral $\int_x^{\infty}\frac{\partial c\left(x_1, t\right)}{\partial t} \cdot d x_1$ were determined numerically from the experimental concentration profile $c(x,t)$. The temperature dependence of the diffusion coefficient $D(c,T)$ can be evaluated by repeating the concentration-dependent diffusion measurements at multiple temperatures. Based on the sample geometry, the furthest distance hydrogen can travel in the sample before interfering with the hydrogen gradient coming in from the other side is $4\,$mm. The boundary condition in Eq. \ref{eq:stenlundcondition} therefore becomes:

\begin{equation}   
\left( \frac{\partial c(x, t)}{\partial x}\right)_{x = 4\,\text{mm}} = 0,
\label{eq:stenlundcondition2}
\end{equation}
This condition places a limit on the maximum time of each measurement. \\

The concentration and temperature dependence of the chemical diffusion coefficient $D_c(c, T)$ is an activated process and can be written as \cite{10.1088/0022-3719/20/10/013, 10.1007/bfb0103398}:
    
\begin{equation}
   D_\text{c}(c, T)=f_\text{therm}(c,T)f_{\text{M}}(c)V(c)D_0,
    \label{eq:Dchem}
\end{equation}
where $D_c$ is the chemical diffusion coefficient, $f_\textrm{M}$ is the mobility correlation factor, $V$ is the site availability factor, $D_0$ is the intrinsic diffusion coefficient:
\begin{equation}
   D_0=\frac{l^2}{6\tau_0}=D_{00}\exp{\frac{-E_\text{act}}{k_\text{B}T}},
    \label{eq:tau0}
\end{equation}
 where <$l$> is the average distance between sites in the glass, $\tau_0$ is the mean time between attempted jumps, $E_\text{act}$ is the activation energy, $k_\text{B}$ is the Boltzmann constant and $T$ is the absolute temperature. The thermodynamic factor $f_{\text{therm}}(c,T)$ is defined as: 
\begin{equation}
f_{\text{therm}}(c,T)=\frac{c}{k_{\text{B}}T}\cdot\frac{\partial\mu}{\partial c},
\label{eq:ftherm}
\end{equation}
where $\mu$ is the chemical potential driving the hydrogen diffusion. $\mu$ can be determined experimentally, for example, by determining the pressure-composition isotherms of the sample at various temperatures. 

The site-availability factor $V(c)$, describes the fact, that there is a chance for every attempted jump, that the approached site is already occupied by another hydrogen. $V(c)$ is defined as the ratio of available to occupied sites around each hydrogen interstitial. There is a distribution of site availability with local hydrogen concentration, where the lower the local hydrogen concentration is, the higher the site-availability factor will be. The mobility correlation factor $f_\text{M}(c)$ describes the change in probability of hydrogen making a correlated backwards jump in the presence of a chemical potential gradient. With hydrogen moving away from an area of increased concentration, there is a higher likelihood of sites being available in front of the hydrogen while sites behind it are more likely to be occupied.

The hydrogen mobility, i.e., motion under the influence of an external force, in this case induced by the chemical potential gradient, is defined as: 
\begin{equation}
    M(c,T) =\frac{D_\text{c}}{c\frac{\partial\mu}{\partial c}}=\frac{D_\text{c}(c,T)}{f_\text{therm}(c,T)\cdot k_\text{B}T}.
    \label{eq:mobility}
\end{equation}

Error bars for $D_\text{c}(c,T)$ were determined at each temperature and concentration as described in \cite{Huang2016}. Each diffusion coefficient reported in this paper is the average of $50$ to $300$ diffusion coefficients determined at that concentration and temperature throughout the measurement time. The error bars provided are the standard deviation of this average. The error bars of the activation energy $E_\text{act}$ were the error of the least-mean-square fit to the diffusion coefficients, weighted with their respective errors.

\section{Results} \label{sec:results}
\subsection{Transmission and concentration profiles}
\begin{figure}
    \centering
    \includegraphics[width = \linewidth]{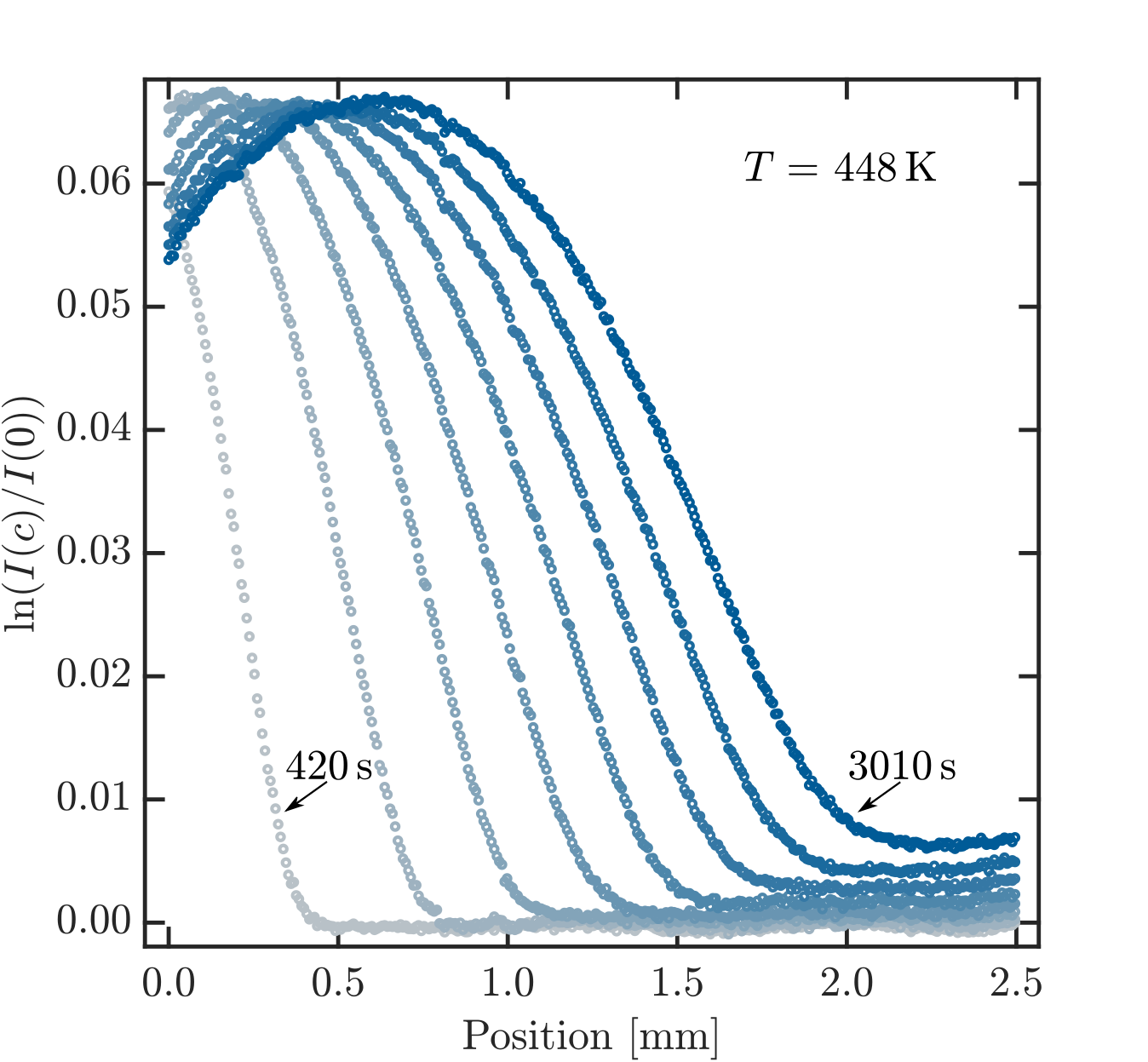}
    \caption{Transmission profile of a V$_{0.8}$Zr$_{0.2}$ metallic glass thin film at $448\,$K. A position of $0\,$mm refers to the edge of the Al$_2$O$_3$ layer, see Fig.~\ref{fig:sampledesign}. More than $50$ individual profiles were evaluated over $3010\,$s. The profiles depicted here are a selection and show the spatial changes in optical transmission every $370\,$s. As the sample is continuously exposed to hydrogen gas, the light transmission through the sample changes systematically, tracing the hydrogen absorption into the sample.}
    \label{fig:trmvx}
\end{figure}

Fig. \ref{fig:trmvx} shows a representative selection of hydrogen-induced changes in the transmission profile measured at $448\,$K. Similar results are obtained at other temperatures, as discussed in Sec. \ref{sec:discussion}. The color intensity of the data points indicates the passage of time. The change in transmitted intensity initially increases before decreasing as the profile evolves. This behavior is rooted in the combination of the evolution of the hydrogen concentration in the sample combined with a parabolic dependence of the transmitted intensity ratio~\cite{Bylin2024}. Since this parabolic dependence has been found and evaluated before, it can be used to determine the underlying concentration profile. The procedure is susceptible to the accuracy and precision of the concentration determination and measurement of the optical transmission but is conceptually straightforward.

\begin{figure}
    \centering
    \includegraphics[width = \linewidth]{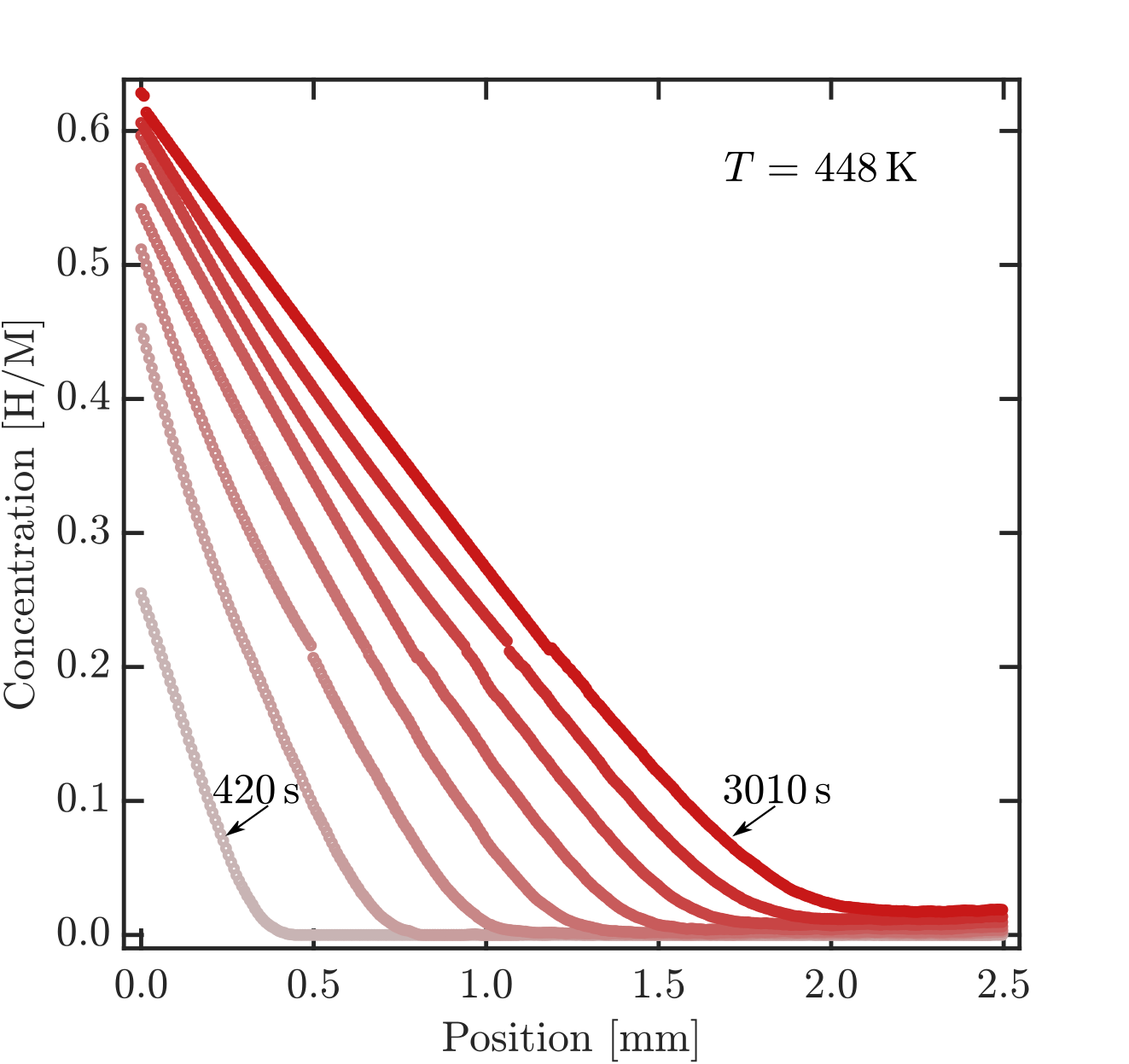}
    \caption{Concentration profile $c(x,t)$ in a V$_{0.8}$Zr$_{0.2}$ metallic glass thin film. The spatial distribution of hydrogen is depicted every $370\,$s, shown from left to right in the plot. The profile was obtained from the optical transmission profile.}
    \label{fig:concvx}
\end{figure}

Fig. \ref{fig:concvx} shows the concentration profile obtained from the optical transmission. The smooth, monotonic nature of the profiles indicates a successful conversion to concentration. The profiles are similar to textbook concentration profiles with a linearly increasing diffusion coefficient as seen in \cite{Crank1975}.

\subsection{Diffusion coefficients and activation energy}
\begin{figure}[ht]
    \centering
    \includegraphics[width = \linewidth]{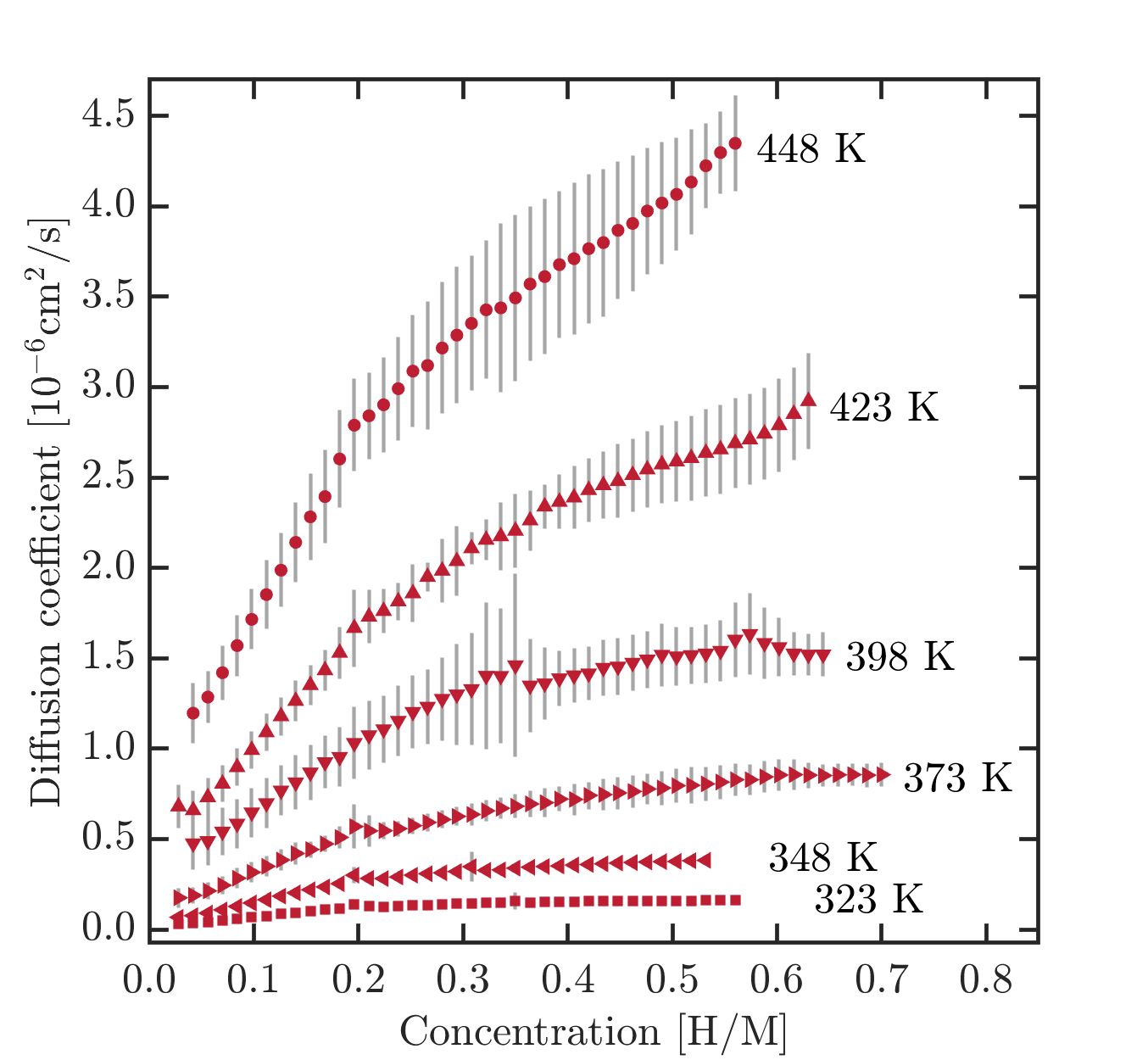}
    \caption{Concentration-dependent diffusion coefficients at different temperatures. The diffusion coefficients were derived from the time-dependent concentration profile $c(x,t)$ at each individual temperature using Eq. \ref{eq:stenlund}.}
    \label{fig:Dvc}
\end{figure}

The concentration and temperature-dependent diffusion coefficient $D_\text{c}(c,T)$, obtained from numerical integration of the concentration profiles using Eq. \ref{eq:stenlund} is shown in Fig. \ref{fig:Dvc}. The chemical diffusion coefficient increases linearly with concentration with a change of slope around $c = 0.2\,$H/M across all temperature ranges. 

\begin{figure}[ht]
    \centering
    \includegraphics[width = \linewidth]{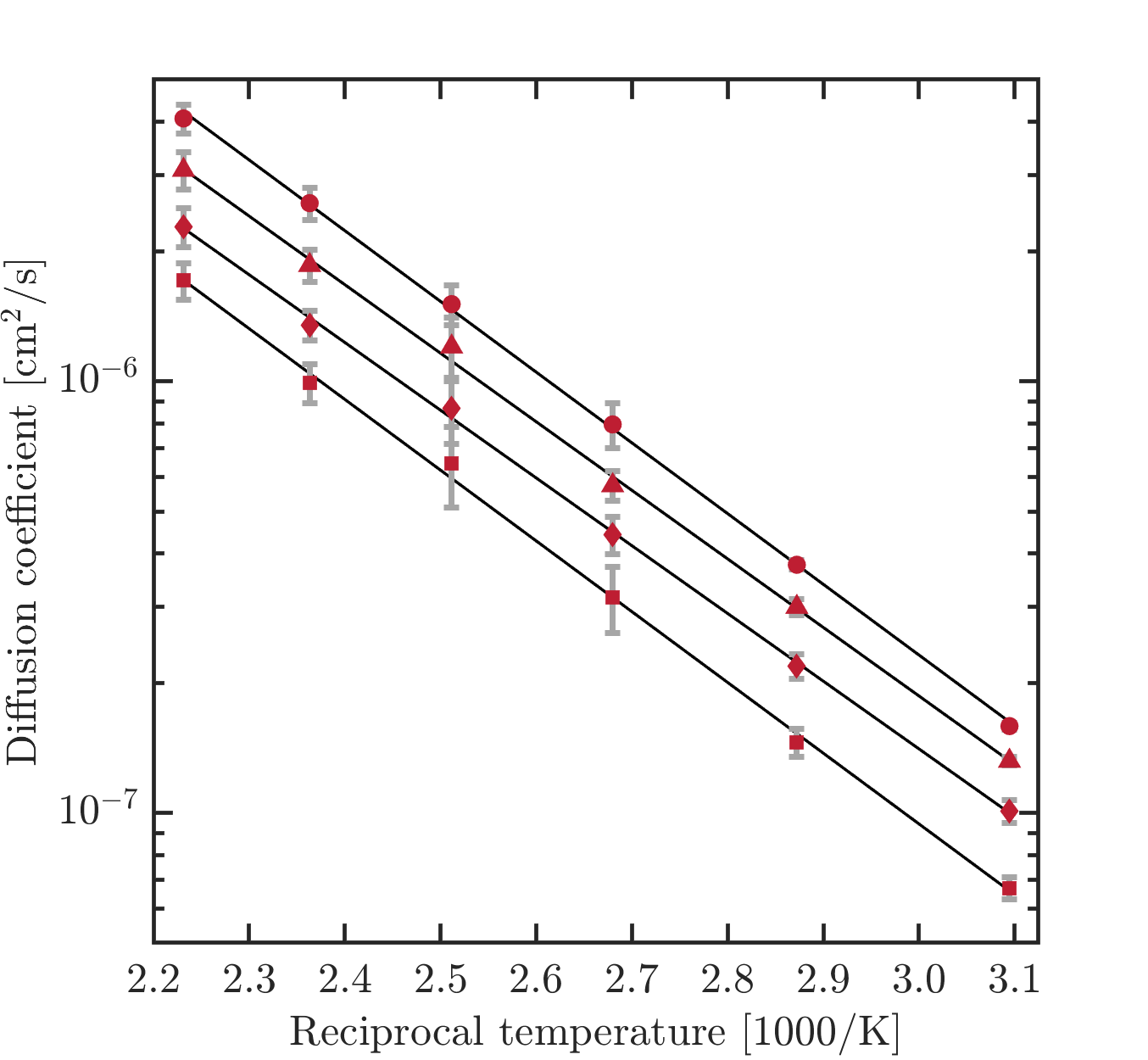}
    \caption{Arrhenius plots of V$_{0.8}$Zr$_{0.2}$ for concentrations of $\{0.10, 0.15, 0.25, 0.50\}\,$H/M, from top to bottom. Good agreement is found between the linear fit of the Arrhenius model (solid black line) and the extracted diffusion coefficients (red symbols).}
    \label{fig:Arrheniusplot}
\end{figure}

\begin{figure}[ht]
    \centering
    \includegraphics[width = \linewidth]{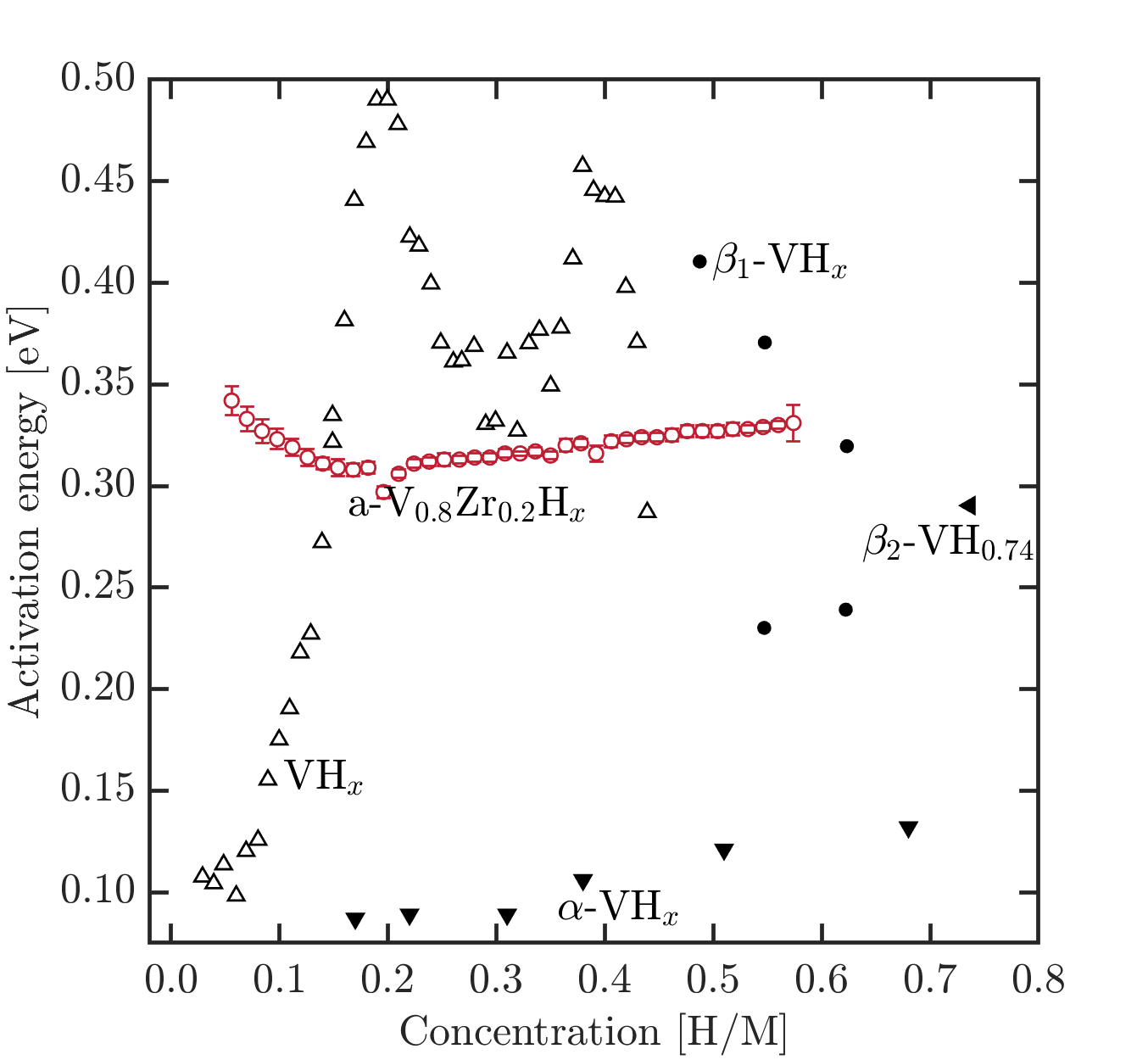}
    \caption{Activation energy $E_\text{act}(c)$ of V$_{0.8}$Zr$_{0.2}$ (open red circles) at varying concentrations and comparison to activation energies of crystalline vanadium thin films \cite{Huang2016} (open black triangles) and bulk vanadium hydride (closed black shapes) \cite{Fukai1977}.}
    \label{fig:Eactvc}
\end{figure}
Fig. \ref{fig:Arrheniusplot} shows the Arrhenius plot of the diffusion coefficients of V$_{0.8}$Zr$_{0.2}$H$_\text{x}$ at concentrations of $\{0.10, 0.15,0.25, 0.50\}\,$H/M. The thermodynamic factor has a temperature dependence that needs to be accounted for prior to an Arrhenius analysis. For that purpose, it was extracted from the isotherms at several different temperatures, (see supplementary Fig S1) and was found to be weakly temperature dependent. It therefore did not significantly influence the concentration dependence of the activation energy. The linear behavior of the data in Fig.~\ref{fig:Arrheniusplot} further supports this argument. We find no evidence of anomalous non-Arrhenius behavior over the studied temperature and concentration range, in stark contrast to measurements of metallic glasses of TiCuH$_{1.41}$ and Zr$_2$PdH$_x$~\cite{Bowman1982,Bowman1983,Bowman1988,Eliaz1999a}. Since a change in the diffusion mechanism is expected to correspond to a change of slope in the Arrhenius plot, we conclude that a change of diffusion mechanism underlying the hydrogen diffusion in V$_{0.8}$Zr$_{0.2}$H$_x$ is unlikely in this temperature and concentration range.

The activation energy $E_\text{act}(c)$ as obtained from the Arrhenius analysis is shown in Fig. \ref{fig:Eactvc} as open red circles. Also shown in the figure is the concentration dependence of the activation energy of hydrogen in epitaxial vanadium films  (open black triangles) \cite{Huang2016}, as well as $\alpha$- and $\beta$-phase poly-crystalline bulk vanadium hydride determined by nuclear magnetic resonance (filled black shapes) \cite{Kleiner1986, Fukai1977}. The activation energy in V$_{0.8}$Zr$_{0.2}$H$_\text{x}$ is a factor of three higher than for hydrogen in the $\alpha$-phase of bulk vanadium and of the same order of magnitude as found in the $\beta$-phases of vanadium hydride.  The slopes of the activation energy of the glass and the disordered $\alpha$-phase appear similar, suggesting a common mechanism. In epitaxial vanadium, on the other hand, the activation energy increases from 0.1 eV at concentrations below 0.1 H/M to 0.47 eV at 0.2 H/M, likely due to a change of site owing to the one-dimensional nature of the allowed expansion. No such behavior is seen in the glass, suggesting no dramatic concentration-dependent change to the depth of the local sites with expansion of the glass.

\begin{figure}
    \centering
    \includegraphics[width=\linewidth]{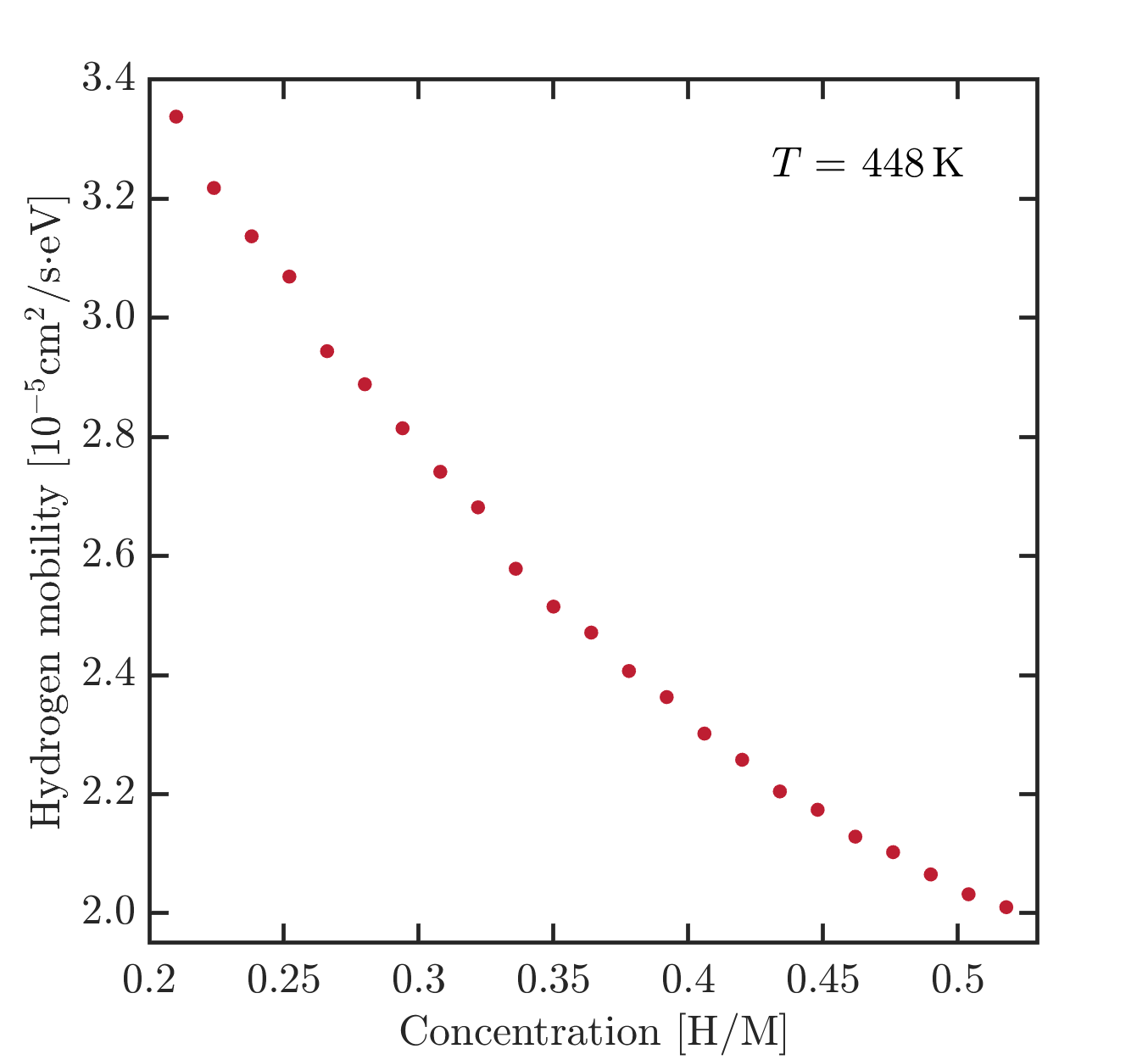}
    \caption{Hydrogen mobility derived from the chemical diffusion coefficient $D_\text{c}$ and the thermodynamic factor $f_\text{therm}$ as explained in the text.}
    \label{fig:mobility}
\end{figure}

In our experiments, we expose the sample to a chemical potential gradient, which modifies the potential barriers, and results in a different quantity obtained than those determined with equilibrium techniques where the thermodynamic factor and the mobility correlation are absent and the tracer correlation factor is present. Since we have estimated the thermodynamic factor, we can determine the mobility $M$ of the hydrogen under this generalized force (the chemical potential gradient) using eq.~{\ref{eq:mobility}}, the results of which are shown in Fig.~\ref{fig:mobility}. This quantity is analogous to the electron mobility under the influence of a potential difference, which in a metallic glass also obeys a diffusion equation \cite{10.1007/978-1-4757-0465-5}. The mobility exhibits a monotonically decreasing behavior over the concentrations shown. Lower concentrations are not shown due to the lack of data for the determination of the thermodynamic factor in this range. The decrease in mobility is attributed to the slight increase in activation energy with concentration. \\

\begin{figure}
    \centering
    \includegraphics[width=\linewidth]{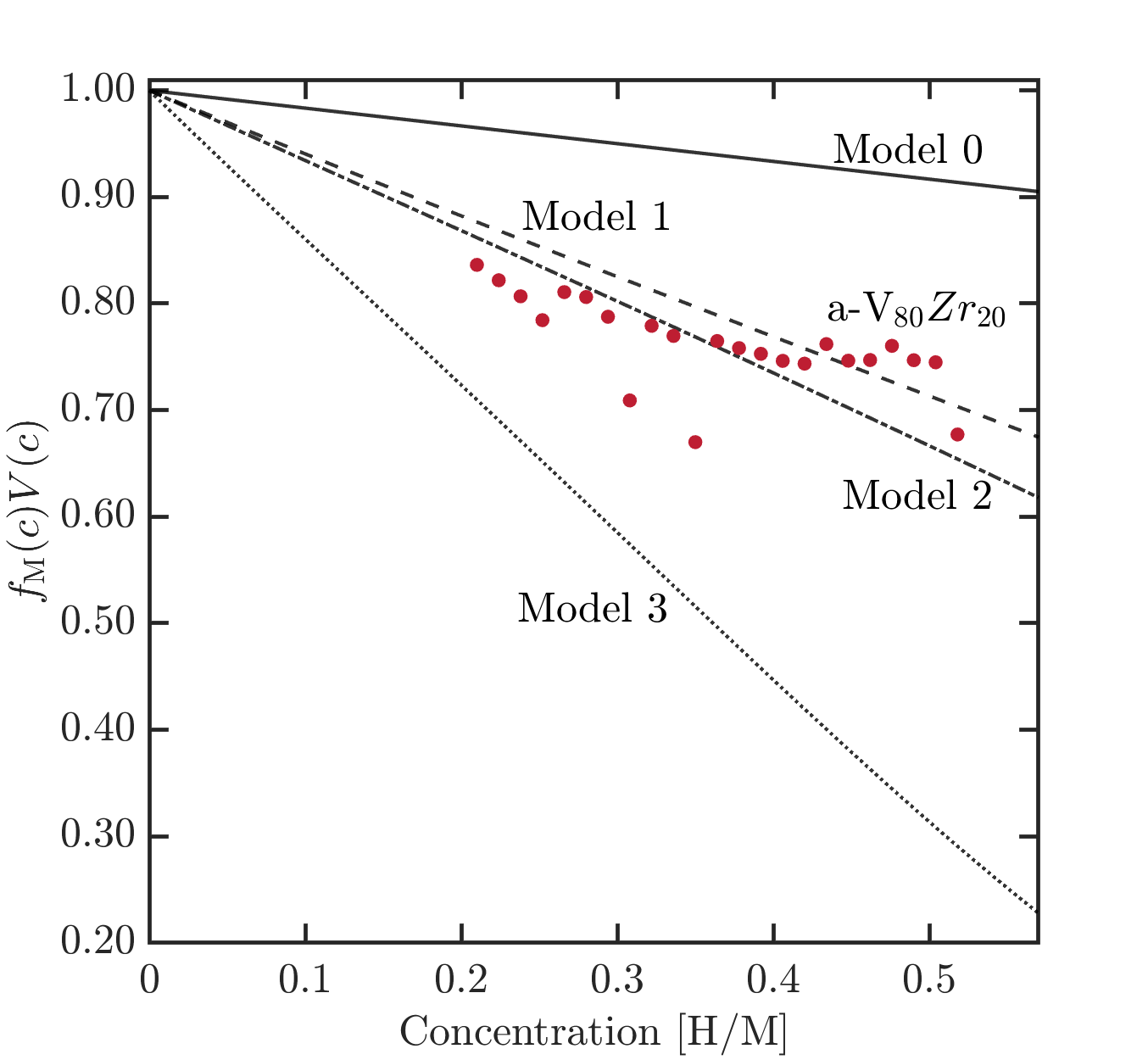}
    \caption{Product of $f_\text{M}(c)V(c)$ for V$_{0.8}$Zr$_{0.2}$ (red circles, $T = 448\,$K) and various models derived in \cite{10.1088/0022-3719/20/10/013}. The data was obtained with an estimate of $D_{00} = 5.51\cdot10^{-3}\,$cm$^2$/s, as explained in the text. The black lines correspond to polynomial fits for different degrees of interstitial site blocking: On-site blocking (solid), $1$st (dashed), $2$nd (dash-dotted) and $3$rd nearest neighbor blocking (dotted).}
    \label{fig:correllation}
\end{figure}

To the extent that the experimental uncertainties allow, a first glimpse into the correlated behavior of hydrogen in the metallic glass, under the influence of a chemical potential gradient, can be obtained by dividing the chemical diffusion by the thermodynamic factor and the exponential. Such a quantity should resemble:

\begin{equation}
   \frac{D_\text{c}}{f_\text{therm}\exp{\frac{-E_\text{act}}{k_\text{B}T}}}=f_{\text{M}}(c)V(c)D_{00},
\end{equation}
and is shown in red in Fig.~\ref{fig:correllation}. The mobility correlation and site availability factor have been determined using Monte Carlo simulations for hydrogen in bcc metals by Faux and Ross~\cite{10.1088/0022-3719/20/10/013} for different models of blocking. It is of interest to compare our results to see if the short-range repulsive nature of the hydrogen-hydrogen interaction bears any resemblance to the one found in the bcc metals. The comparison with bcc metals is meaningful since the glass is composed of $80$\% vanadium, which is bcc in its ground state. The estimate for $D_{00}$ was determined ($D_{00} = 5.51\cdot10^{-3}\,$cm$^2/$s) by comparing our data to the models of Faux and Ross. We treat this as the only free parameter in this paper, and with this caveat can see that the concentration dependence resembles models 1 or 2 proposed by Faux and Ross. These two models correspond to nearest neighbor blocking and second nearest blocking, respectively. Faux and Ross found that model 2 most resembles the experimental case in the disordered $\alpha$-phase in bcc tantalum. These findings are therefore physically plausible and indicate that the repulsion in the glass is similar to the crystalline case and, to first approximation, does not depend on the crystal structure. We call for a theoretical treatment of the mobility correlation and site availability factor for metallic glasses to further assess the veracity of this statement. 

\section{Discussion} \label{sec:discussion}
To our knowledge, this is the first time the diffusion of hydrogen has been evaluated in a metallic glass over such a wide range of concentrations. At first sight the increase in the chemical diffusion coefficient with concentration is in stark contrast to the same quantity in bulk vanadium or epitaxial vanadium films. However, close examination of Eq. \ref{eq:Dchem} and the data provided in Sec. \ref{sec:results}, one can explain the monotonic increase of $D_\text{c}(c,T)$ with concentration shown in Fig. \ref{fig:Dvc}. The product $f_\text{M}(c)V(c)$ monotonically decreases over the measured concentration range. It can therefore not drive the monotonic increase in diffusion with concentration reported in Fig. \ref{fig:Dvc}. 
The activation energy shows a slight increase with concentration above $c=0.11$ H/M (see Fig.~\ref{fig:Eactvc}), which contributes to a minor decrease of the diffusion coefficient. The remaining quantity, the thermodynamic factor shown in the supplementary Fig. S3, increases with the hydrogen concentration in the measured range. This is in contrast to the thermodynamic factor in crystalline vanadium that goes down monotonically from $0.00$H/M to $0.35\,$H/M and then increases owing to the large changes in slope of the pressure-composition isotherms, which is generally seen in crystalline materials. In a metallic glass, over the range that the isotherms can be approximated by linear functions, one can therefore expect an increase in diffusion due to the thermodynamic factor. The thermodynamic factor therefore improves the speed of hydrogen in glasses (in the linear range) whereas it hinders it in crystalline materials. This fact is also reflected in the mobility shown in Fig.~\ref{fig:mobility}, primarily going down due to the intrinsically activated nature of the diffusion process and not due to the thermodynamic factor.

It has been shown that clamping of thin films (adhesion to the substrate) significantly affects the hydrogen diffusion coefficients in crystalline VH$_{x}$ \cite{Huang2016}. In crystalline vanadium, hydrogen is known to occupy specific interstitial sites within the crystal lattice. Located on either a tetrahedral or an octahedral site, hydrogen expands the atomic lattice around it, leading to local and global lattice deformations and ultimately a phase change of the material from a solid solution to the hydride. In contrast to the bulk case, where the sample can expand freely in all three spatial dimensions, the thin film sample can only expand freely in one dimension: away from the sample substrate. In crystalline vanadium, this so called clamping of the sample material effectively limits the number of 
sites that can freely be occupied by the hydrogen, leading to an overall reduction in diffusion coefficients \cite{Huang2016}. In metallic glasses, the sites are distributed more homogeneously throughout the solid. The thin film form is still likely to lead to a reduction in energetically favorable sites as the sample is clamped just as the crystalline one. 
The activation energy reported for the V$_{0.8}$Zr$_{0.2}$ metallic glass has a similar concentration dependence as the $\alpha$-phase of bulk vanadium, which suggests that the topological disorder present in the glass is not the driving factor for the increase in activation energy. This was argued to be the case by Driesen and Kehr for hydrogen diffusion in general \cite{10.1103/physrevb.39.8132} and is further supported here. 
From a broader perspective, a low concentration dependence of the activation energy could be advantageous for applications regarding the distribution of hydrogen throughout the system. Since hydrogen diffusion is faster as concentration increases, local accumulation of hydrogen within the material is inhibited. These local accumulations result in large local stress fields, which can lead to deformation and eventually hydrogen embrittlement. As metallic glasses do not exhibit grain boundaries, nor hydride phases of different specific volumes, having a uniform diffusion coefficient means consistent delivery of hydrogen in and out of the material. In MgH$_2$ for example, the diffusion in the hydride phase is prohibitively slow and is the limiting factor for hydrogen uptake and release. This problem is completely avoided in materials with a flat activation energy profile with hydrogen concentration. \\

\section{Conclusions} \label{sec:conclusions}
The optical transmission method can be used to determine the concentration-dependent diffusion coefficient $D_\text{c}(c,T)$ and activation energy $E_\text{act}(c)$ of metallic glass thin films, as long as they are optically responsive to the absorption of hydrogen. The results shown here are the first-ever concentration-dependent diffusion measurements of metallic glasses over a range of concentration as large as $0.0\,\text{to}\,0.6\,$H/M. The diffusion behavior found in thin metallic glass films of V$_{0.8}$Zr$_{0.2}$ is in stark contrast to that of crystalline VH$_x$-films since in the glass it initially increases with concentration but decreases in the crystalline phase. The difference is attributed to the thermodynamic factor being radically different in crystalline materials and metallic glasses. However, more concentration-dependent measurements of diffusion coefficients and activation energies in other metallic glasses are needed to gain further insights as to the generality of this trend. The straightforward nature of simple light transmission measurements makes the method an attractive choice for studying hydrogen motion in metallic glasses. By increasing the pressure of the hydrogen atmosphere, the concentration range can be extended up to the maximum concentration possible in the sample. This measurement platform together with the thin film design allows the exploration of the physics of hydrogen in low dimensional structures, such as changes of elastic boundary conditions, that lead to clamping, uniaxial volume changes, and finite-size effects.
The method paves the way for combinatorial screening of H diffusion in metallic glasses, whereby multiple compositions can be measured simultaneously without the need for moving parts or exchange of samples.\\

\section*{Acknowledgements} \label{sec:acknowledgements}
The authors are indebted to Dr. J. Bylin for fruitful discussions and for providing data on the isotherms. They also gratefully acknowledge the Swedish Research Council (VR -  Vetenskapsrådet) grant number 2018-05200 for funding the work and the Swedish Energy Agency grant 2020-005212 (50697-1). G.K.P. furthermore acknowledges the Carl-Tryggers Foundation grants CTS 17:350, CTS 19:272, and CTS 22:2091 for financial support.

\bibliography{Library_for_the_Diffusion_Paper.bib}

\end{document}